\documentclass[journal]{IEEEtran}
\usepackage[noadjust]{cite}
\usepackage{amsmath,amssymb,amsfonts}
\usepackage[bookmarks=false]{hyperref}
\hypersetup{draft}

\usepackage{graphicx}
\usepackage{textcomp}
\usepackage{xcolor}
\newcommand{\etal}{\textit{et al. }}
\usepackage{caption}
\usepackage{amsmath}
\usepackage{amssymb}
\usepackage{graphicx}
\usepackage{enumitem}

\usepackage{algorithmicx}
\usepackage[ruled,linesnumbered]{algorithm2e}
\usepackage{algpseudocode}
\usepackage{multirow}
\usepackage{graphicx}
\usepackage{subfigure}
\usepackage{booktabs}
\begin{document}

\title{Exploiting Vulnerabilities of Deep Learning-based Energy Theft Detection in AMI through Adversarial Attacks}

\author{
    \IEEEauthorblockN{Jiangnan Li$^{\ast}$, Yingyuan Yang$^{\S}$, Jinyuan Stella Sun$^{\ast}$} \\
    \IEEEauthorblockA{$^{\ast}$The University of Tennessee, Knoxville, {\{jli103, jysun\}@utk.edu}
    \\$^{\S}$University of Illinois Springfield, yyang260@uis.edu}
}

\maketitle

\begin{abstract}
Effective detection of energy theft can prevent revenue losses of utility companies and is also important for smart grid security. In recent years, enabled by the massive fine-grained smart meter data, deep learning (DL) approaches are becoming popular in the literature to detect energy theft in the advanced metering infrastructure (AMI). However, as neural networks are shown to be vulnerable to adversarial examples, the security of the DL models is of concern. 

In this work, we study the vulnerabilities of DL-based energy theft detection through adversarial attacks, including single-step attacks and iterative attacks. From the attacker's point of view, we design the \textit{SearchFromFree} framework that consists of 1) a randomly adversarial measurement initialization approach to maximize the stolen profit and 2) a step-size searching scheme to increase the performance of black-box iterative attacks. The evaluation based on three types of neural networks shows that the adversarial attacker can report extremely low consumption measurements to the utility without being detected by the DL models. We finally discuss the potential defense mechanisms against adversarial attacks in energy theft detection.

\end{abstract}


\ifCLASSOPTIONpeerreview
\begin{center} \bfseries EDICS Category: 3-BBND \end{center}
\fi

\IEEEpeerreviewmaketitle

\section{Introduction}

\IEEEPARstart{E}{nergy} theft is one of the primary non-technical losses (NTL) in the power systems and causes high financial losses to electric utility companies around the world \cite{gao2019physically}. In recent years, two-way data communications between the customers and utilities are enabled by the development of the advanced metering infrastructure (AMI). Smart meters that provide fine-grained power consumption data of customers are expected to mitigate energy theft. However, AMI introduces new vulnerabilities that can be exploited by the attackers to launch energy theft \cite{zanetti2017tunable}. The smart meters are shown to be vulnerable to physical penetration \cite{hackmeter} and there are even video tutorials online on smart meter hacking \cite{youtubeHacking}. To date, the energy theft problem is still serious and the corresponding detection approaches are needed.

Currently, the energy theft detection approaches proposed in the literature can be categorized into two classes, sensor-based detection and user profile-based. The sensor-based detection requires extra equipment to be deployed in AMI while the user profile-based detection exploits the customer's power usage patterns to detect abnormal variations. In recent years, machine learning (ML) techniques, especially deep learning (DL), are studied in the literature to detect energy theft \cite{zheng2018novel, ismail2020deep, zanetti2017tunable, gao2019physically, nabil2018deep, jokar2015electricity, punmiya2019energy, zheng2017wide, hasan2019electricity, korba2018energy}. The ML-based detection approaches take advantage of the massive fine-grained power consumption data of smart meters and can achieve state-of-the-art performance. Meanwhile, ML approaches are usually pure software systems, which make them compatible with the legacy power system infrastructures.

However, recent studies in the computer vision (CV) domain have shown that the well-trained DL models are highly vulnerable to adversarial examples \cite{szegedy2013intriguing, goodfellow2014explaining, rozsa2016adversarial, moosavi2016deepfool, carlini2017towards}. By adding a well-crafted perturbation vector to the legitimate input, adversarial attackers can deceive the DL models to output wrong prediction results. In addition, the same perturbation is transferable between different DL models that own different structures and are trained with different datasets. Adversarial attacks are also demonstrated to be effective in power system applications \cite{chen2018machine, chen2019exploiting, li2020conaml}. As DL becomes a popular technique in energy theft detection, the potential risks of adversarial attacks need to be investigated, which is the focus of this paper.

Although sophisticated adversarial machine learning (AML) algorithms are proposed in the CV domain, they can not be applied to attack DL-based energy theft detection directly due to different requirements for the examples. For instance, the adversarial perturbations in the CV domain are required to be small so that the adversarial image will be hardly perceptible to human eyes. Since such constraints do not apply to an energy thief, the performances of the corresponding AML algorithms in energy theft detection need to be evaluated. Meanwhile, to increase the stolen profit, the energy thief should focus on the size of the adversarial example (power consumption measurements reported to the utilities) instead of the divergence between original false measurements and adversarial measurements. Therefore, the adversarial attack that maximizes the attacker's profit should be formulated to evaluate the reliability of the DL detection models.

In this paper, we investigate the vulnerabilities of DL-based energy theft detection through AML, including single-step attacks and iterative attacks. We design \textit{SearchFromFree}, a framework to increase the attacker's profit. The main contributions of this paper, which is an extension of a shorter conference paper \cite{li2020searchfromfree}, can be summarized as follows:

\begin{itemize}
\item We study the vulnerabilities of DL-based energy theft detection and summarize the properties of adversarial attacks in energy theft detection by proposing a general threat model.

\item We propose a random adversarial measurement initialization approach that maximizes the attacker's profit. The initialization approach is compatible with different AML algorithms and can generate valid adversarial examples with low energy costs. Meanwhile, based on the threat model, we design an iterative adversarial measurement generation algorithm that employs a step-size searching scheme to increase the performance of black-box attacks.

\item The evaluations are implemented with three types of neural networks that are trained with a real-world smart meter dataset, and the result demonstrates that our algorithm can generate extremely small false power consumption measurements that can still successfully bypass the DL models' detection. 

\item We discuss the defense mechanisms to increase the reliability of DL detection methods.

\end{itemize}

Related research is discussed in section \ref{sec:related}. Section \ref{sec:formation} formats the problem, proposes the threat model, and presents the design of algorithms. \ref{sec:evaluation} presents the experiment evaluation. The defense mechanisms are discussed in section \ref{sec:defense}. Finally, section \ref{sec:conclusion} concludes the paper.

\section{Related Work} \label{sec:related}

The support vector machine (SVM) is employed by Nagi \etal to detect abnormal power usage behaviors based on the historian consumption data in 2009 \cite{nagi2009nontechnical}. After that, Depuru \etal extended their approach and included more features, such as the type of consumer, geographic location, to increase the performance of the SVM classifier \cite{depuru2011support}. In 2015, Jokar \etal generated a synthetic attack dataset and trained a multiclass SVM classifier for each customer to detect malicious power consumption \cite{jokar2015electricity}. Besides, SVM can also be combined with other techniques, such as a fuzzy inference system \cite{nagi2011improving} and decision tree \cite{jindal2016decision}. In 2017, Zheng \etal employed deep convolutional neural networks (CNN) to detect energy theft based on a real-world dataset and achieved a high detection rate \cite{zheng2017wide}. After that, \cite{nabil2018deep} trained a deep recurrent neural network (RNN) with the Irish smart meter dataset \cite{IrishData} and randomly searched for model parameters. In 2020, Ismail \etal studied energy theft from the distributed generation domain and proposed a hybrid neural network detection model \cite{ismail2020deep}. Other DL-based energy theft detection approaches can be found in \cite{hasan2019electricity, korba2018energy}.

In 2013, Szegedy \etal firstly proposed the adversarial attacks to deep neural networks \cite{szegedy2013intriguing} in the CV domain. After that, various AML algorithms were proposed, such as the Fast Gradient Sign Method (FGSM) by \etal \cite{goodfellow2014explaining}, Fast Gradient Value (FGV) by Rozsa \etal \cite{rozsa2016adversarial}, and DeepFool by Moosavi-Dezfooli \etal \cite{moosavi2016deepfool}. Recently, the impact of adversarial attacks on power system ML applications is also investigated. In 2018, Chen \etal showed that both the classification and regression applications in the power system are vulnerable to adversarial attacks. They then launched adversarial attacks to study the vulnerabilities of load forecasting \cite{chen2019exploiting}. In 2019, Liu \etal showed that the ML-based AC state estimation can be compromised by adversarial attacks \cite{liu2019adversarial}. \cite{li2020conaml} studied the DL-based false data injection attack and demonstrated that the adversarial attacker can compromise both the DL detection and residual-based detection under physical constraints.  Marulli \etal studied the data poisoning attacks to ML models in energy theft detection using the generative adversarial network (GAN) \cite{marulli2019adversarial}. However, they did not consider the evasion attacks and the profit of the attacker . 

\section{Formation and Design} \label{sec:formation}

\subsection{Adversarial Energy Theft Formation}

To launch the energy theft, the attacker is assumed to be able to compromise his/her smart meter and can freely modify the power consumption measurements that are reported to the utilities. In general, the DL-based energy theft detection can be considered as a binary classification problem. Given the power consumption measurements $M$, the utility company utilizes a DL classifier $f_{\theta}: M \rightarrow Y$ trained by a dataset $\left \{ M, Y \right \}$ to map the measurements $M$ to their labels $Y$ (Normal or Theft). Therefore, the adversarial attack in energy theft detection should a false-negative attack that deceives the DL classifier $f_{\theta}$ to categorize the adversarial measurement vectors $A$ as normal. Meanwhile, $A$ needs to be small so that the energy thief can obtain a high profit. Without loss of generality, the adversarial attack in energy theft can be represented as an optimization problem:

\begin{subequations}
	\begin{align}
	\min \;\;\;& \left \| a \right \|_{1}\\
	s.t. \;\;\; & f_{\theta}(a) \rightarrow Normal \label{eq:pass}\\ 
	& a_{i} \geq 0 \label{eq:bepositive}
	\end{align}
\label{eq:formation}
\end{subequations}

where $a$ represents a specific adversarial measurement vector, $\left \| a \right \|_{1} = \sum a_{i} $ is the $L_1$-Norm of $a$. The constraint (\ref{eq:bepositive}) requires that all the power consumption measurement $a_{i}$ in $a$ must be non-negative to be feasible.

\subsection{Threat Model}
\label{sec:threatmodel}

We propose a practical threat model for the adversarial attacks in energy theft detection, as described below:

\begin{itemize}

\item The attacker can compromise his/her smart meter and freely modify the meter's power consumption measurements reported to the utilities. In practice, this can be implemented through physical penetration.

\item If a DL model was trained by the utilities, it will usually be deployed on a separate server that owns isolated access networks.  Therefore, we consider a black-box adversarial attack that the energy thief have no access to the utilities' DL model $f_{\theta}$ and training dataset $\left \{ M, Y \right \}$.

\item We allow the attacker to obtain an alternative dataset $\left \{ M', Y' \right \}$, such as a historian or public dataset, to train his/her model $f'_{\theta'}$ to generate adversarial measurements. The principle behinds the black-box attack is the transferability of adversarial examples.

\item As demonstrated by (\ref{eq:bepositive}), the attacker needs to generate non-negative adversarial measurements to be practical.

\end{itemize}

In addition to the black-box attack described above, we will also evaluate the performance of white-box attacks that allows the attacker to fully access the DL model $f_{\theta}$, such as insider attackers. Such evaluations can study the reliability of the detection system under the worst-case scenario and the upper bound performance of the adversarial attacks.

\subsection{State-of-the-art Approaches} \label{sec:sotap}

Since the constraint (\ref{eq:pass}) defined by the neural network is highly-nonlinear, formation (\ref{eq:formation}) is difficult to solve directly by existing optimization approaches. Generally, the existing AML algorithms maximize the adversarial attack performance by increasing the classification loss of the DL model through gradient-based optimization. We release the related constraints, such as the box-constraint, in the CV applications so that they can fit the energy theft attack requirements.

In general, the AML algorithms can be categorized into single-step attacks and iterative (multiple-step) attacks. The single-step attacks usually have better transferability but are relatively easy to defend, while the iterative attacks are more powerful but are less transferable \cite{kurakin2016adversarial}. In this paper, we study three state-of-the-art AML algorithms, FGSM \cite{goodfellow2014explaining}, FGV \cite{rozsa2016adversarial} and DeepFool \cite{moosavi2016deepfool}, as shown below:

\subsubsection{Fast Gradient Sign Method (FGSM)} 

The FGSM method proposed in \cite{goodfellow2014explaining} is a single-step attack method. Given $a$, FGSM updates the vector according to equation (\ref{eq:fgsm}), where $\epsilon$ is a constant, $L$ is a loss function, and $Y_{a}$ is the label (theft) of $a$. With the $sign$ function, the perturbation vector size is controlled by $\epsilon$.

\begin{equation}
a = a + \epsilon \cdot sign (\nabla_{a} L(f_{\theta}(a), Y_{a}))
\label{eq:fgsm}
\end{equation}

\subsubsection{Fast Gradient Value (FGV)} 

The FGV method proposed in \cite{rozsa2016adversarial} is also a single-step algorithm and is similar to FGSM. However, FGV employs the original gradient values instead of the sign value, as shown in equation (\ref{eq:fgv}).

\begin{equation}
a = a + \epsilon \cdot \nabla_{a} L(f_{\theta}(a), Y_{a})
\label{eq:fgv}
\end{equation}

\subsubsection{DeepFool}

The DeepFool algorithm proposed in \cite{moosavi2016deepfool} is an iterative algorithm that aims to minimize the perturbation size. Assuming the neural networks utilize Softmax as the activation function in the last layer, DeepFool keeps executing equation (\ref{eq:deepfool}) until $a$ can be classified as Normal by $f_\theta$.

\begin{equation}
a = a - \frac{f_{\theta}(a)}{\left \| \nabla_{a}f_{\theta}(a) \right \|_{2}^{2}} \nabla_{a}f_{\theta}(a)
\label{eq:deepfool}
\end{equation}

\subsection{SearchFromFree Framework}

\subsubsection{Randomly Initialization}

The state-of-the-art AML algorithms are originally designed for CV applications where the main constraint is the magnitude of the adversarial perturbation. However, as demonstrated in equation (\ref{eq:formation}) the purpose of the attacker is to minimize the adversarial power consumption measurements reported to the utility to reduce his/her cost. 

Different from CV applications where the input images are given and static, the energy thief can freely modify the smart meter's measurements. Since the AML algorithms can constrain the adversarial perturbation to be small, it is intuitive that the crafted adversarial measurements will be small if the initial measurements for the DL model are small. In practice, the minimum power consumption should be zero, which indicates a free electricity bill to the energy thief. However, a constant zero measurement vector will result in constant adversarial measurements, which is obviously abnormal to the utilities. Therefore, we propose a scheme that randomly initializes adversarial measurements according to a Gaussian distribution $a \sim \mathcal{N}(0, \sigma^{2})$ with the mean value set to zero ($\mu = 0$) and the standard deviation $\sigma$ set to a small number. We set all the non-negative values in $a$ to zero to meet constraint (\ref{eq:bepositive}). This initialization approach is compatible with different AML algorithms.

\subsubsection{Step-size Searching Scheme}

The iterative attacks usually have worse transferability. For example, the multiple-step DeepFool attack executes an iteration process and return the adversarial example as soon as it is misclassified by the given model. Empirically, the example is unique to the given model and may have low transferability. In energy theft detection, the adversarial measurements from the attacker are always smaller than normal measurements. Statistically, for a trained model, larger adversarial measurement vectors will have a higher probability to bypass the detection. Since the attacker's model $f'_\theta$ may share a similar manifold with $f_\theta$, we design a step-size iterative scheme to search adversarial measurements that share the best transferability to increase the performance of black-box iterative attacks.

Enabled by the randomly initialization approach, our step-size scheme can be represented by Algorithm \ref{al:ssf}.

\begin{algorithm}
\SetAlgoLined
\textbf{Input:} $f' _{\theta'}$, $step$, $size$, $\sigma$ \\
\textbf{Output:} $a$\\

\SetKwBlock{Begin}{function}{end function}
\Begin($\text{ssf-Iter} {(} f' _{\theta'}, step, size, \sigma {)}$)
{
  initialize $a \sim N(0, \sigma^{2} )$ \\
  a = \textbf{clip}(a, min=0) \\
  initialize $stepNum = 0$ \\
  
  \While{$stepNum \leq step - 1$ }{
  
  calculate gradient $ G = \nabla_{a} L(f'_{\theta'}(a), Y_{a})$\\

  $r = G \cdot size / \textbf{max}(\textbf{abs}(G))$ \\
  update $a = a + r$ \\
  a = \textbf{clip}(a, min=0) \\
  $stepNum++$ \\
 }
 \Return{$a$}
}
\caption{SearchFromFree Iteration Algorithm}
\label{al:ssf}
\end{algorithm}

The \textbf{ssf-Iter} function in Algorithm \ref{al:ssf} has four inputs, including the local ML model $f'_{\theta'}$ and three positive constant parameters. The constant $step$ limits the maximum number of search iteration while $size$ defines the maximum modification of $a$ in each iteration. As shown by Line 4, we empirically initialize $a$ according to a Gaussian distribution with the standard deviation value equals to $\sigma$ and mean value equals to zero. The iteration process gradually increases $\left \| a \right \|_{1}$ to have a higher probability to bypass the detection. Therefore, a small initial $a$ will finally lead to a smaller $\left \| a \right \|_{1}$ so that the attacker can make more profit. The perturbation $r$ generated from the loss gradient may cause negative measurement values in $a$. To solve this, we set all the negative values to zero to generate a feasible adversarial measurement vector $a$, as shown by Line 5 and 11.

\section{Simulation Evaluation} \label{sec:evaluation}

\subsection{Dataset Structure}

We employ the smart meter data published by the Irish Social Science Data Archive (ISSDA) \cite{IrishData} as it is widely used as a benchmark for energy theft detection in related literature \cite{zheng2018novel, zanetti2017tunable, nabil2018deep, jokar2015electricity, punmiya2019energy, korba2018energy}. The dataset contains the smart meter energy consumption measurement data of over 5000 customers in the Irish during 2009 and 2010. We assume all the measurement data in the dataset is normal since the customers agreed to install the smart meters and participated in the research project. There are missing and illegal measurements in the original dataset and we pre-process the raw dataset by filtering out the incomplete measurements. We then regulate the time-series measurement data into daily reading vectors and obtain the dataset $D_{daily}$. Since the smart meters are recorded every 30 minutes, each daily reading measurement vector will contain 48 power consumption measurements.

To solve the shortage of real-world energy theft datasets, we employ the false measurement data generation approach proposed in \cite{zanetti2017tunable} to simulate the energy theft measurements, which is a benchmark method used in previous literature \cite{zheng2018novel, gao2019physically, nabil2018deep, jokar2015electricity, punmiya2019energy}. \cite{zanetti2017tunable} presents six energy theft scenarios, as shown in Table \ref{table:scenario}. The attack $h_{1}$ multiples the original meter reading with a constant while $h_{2}$ with a random constant generated from a uniform distribution. The $h_{3}$ considers that the energy thief reports zero consumption during a specific period. The $h_{4}$ scenario happens when an attacker constantly reports the mean consumption. $h_{5}$ is similar to $h_{2}$ but multiplying the random constant with the mean value instead of the real measurements. Finally, $h_{6}$ reverses the records of a day so that the small measurements will be reported during the periods in which the electricity price is lower.

\begin{table}[htbp]
\caption{Energy Theft Attack Scenarios \cite{zanetti2017tunable}}
\begin{center}
\begin{tabular}{c}
\toprule[2pt]
\textbf{Attack Scenario}  \\
\midrule[1pt]
$h_{1}(m_{t}) = \alpha m_{t}$, $\alpha \sim Uniform(0.1, 0.8 )$ \\
\hline
$h_{2}(m_{t}) = \beta_{t} m_{t}$, $\beta_{t} \sim Uniform(0.1, 0.8 )$  \\
\hline
$h_{3}(m_{t}) = \left\{\begin{matrix}
0     & \forall t \in \left [ t_{i}, t_{f} \right ]\\ 
m_{t} & \forall t \notin \left [ t_{i}, t_{f} \right ]
\end{matrix}\right.$  \\
\hline
$h_{4}(m_{t}) = E(m)$  \\
\hline
$h_{5}(m_{t}) = \beta_{t} E(m)$ \\
\hline
$h_{6}(m_{t}) = m_{48-t}$\\
\bottomrule[2pt]
\end{tabular}
\end{center}
\label{table:scenario}
\end{table}

A synthetic dataset is generated based on the regulated daily smart meter dataset. We randomly sample 180,000 daily records from $D_{daily}$ and pollute half records in the sampled dataset according to the attack scenarios described in Table \ref{table:scenario}. We label all normal records as 0 and polluted records as 1 and employ One-hot encoding for the labels. We finally obtain a defender's dataset $D_{defender}:\left \{ M_{180,000 \times 48}, Y_{180,000 \times 1} \right \}$. We simulate the dataset $D_{attacker}$ for the attacker in the same way. 

\subsection{Model Training}

\begin{table}[htbp]
\caption{Model Performance}
\begin{center}
\begin{tabular}{c|c|c}
\toprule[2pt]
\textbf{Model} & \textbf{Accuracy} & \textbf{False Positive Rate} \\
\midrule[1pt]
$f_{FNN}$ & 86.9\% & 10.01\% \\
\hline
$f'_{FNN}$ & 86.87\% & 14.01\% \\
\hline
$f_{RNN}$ & 97.5\% & 2.58\% \\
\hline
$f'_{RNN}$ & 97.48\% & 2.62\% \\
\hline
$f_{CNN}$ & 93.49\% & 7.79\% \\
\hline
$f'_{CNN}$ & 93.28\% & 6.41\% \\
\bottomrule[2pt]
\end{tabular}
\end{center}
\label{table:NNperf}
\end{table}

The evaluation experiments are conducted based on three types of deep neural networks (DNN), feed-forward neural network (FNN), CNN, and RNN. We train three DL models for the defender (utilities) and three separate models for the attacker with $D_{defender}$ and $D_{attacker}$ respectively. For each model, 20\% records in $D_{defender}$ or $D_{attacker}$ are randomly sampled for testing the rest 80\% for training. We manually tuned the parameters of the model training and the performances of corresponding models are shown in Table \ref{table:NNperf}. Overall, the RNNs achieve the best classification performance since they have an intrinsic advantage in learning the pattern of time-series data. The structures of the neural networks are shown in Table \ref{table:struct}. All the DNNs are implemented with the TensorFlow and Keras library. The training process is conducted on a Windows 10 PC with an Intel Core i7 CPU, 16 GB memory, and an NVIDIA GeForce GTX 1070 graphic card to accelerate the training process. The models are optimized with a Rmsprop optimizer.

\begin{table*}[htbp]
\caption{Model Structures}
\begin{center}
\begin{tabular}{c|c|c|c|c|c|c}
\toprule[2pt]
\textbf{Networks} & \multicolumn{2}{|c|}{\textbf{FNN}} & \multicolumn{2}{|c|}{\textbf{RNN}} & \multicolumn{2}{|c}{\textbf{CNN}} \\
\midrule[1pt]
\textbf{Models} & $f_{FNN}$ & $f'_{FNN}$ & $f_{RNN}$ & $f'_{RNN}$ & $f_{CNN}$ & $f'_{CNN}$  \\
\hline
\textbf{Layer 0} & input $48$ & input $48$ & input $48 \times 1$ & input $48 \times 1$ & input reshaped $6 \times 8$ & input  reshaped $6 \times 8$  \\
\hline
\textbf{Layer 1} & 128 Dense & 168 Dense & 256 LSTM & 246 LSTM & 128 Conv2D & 156 Conv2D \\
\hline
\textbf{Layer 2} & 256 Dense & 328 Dense & Dropout 0.25 & Dropout 0.25 & 128 Conv2D & 214 Conv2D \\
\hline
\textbf{Layer 3} & 128 Dense & 168 Dense & 168 LSTM & 148 LSTM & MaxPooling2D (2,2) & MaxPooling2D (2,2) \\
\hline
\textbf{Layer 4} & 64 Dense & 128 Dense & Dropout 0.25 & Dropout 0.25 & Dropput 0.25 & Dropput 0.25\\
\hline
\textbf{Layer 5} & Dropout 0.25 & Dropout 0.25 & 128 LSTM & 108 LSTM & flatten & flatten\\
\hline
\textbf{Layer 6} & 32 Dense & 64 Dense & 2 Dense Softmax & 2 Dense Softmax & 32 Dense & 48 Dense \\
\hline
\textbf{Layer 7} & Dropout 0.25 & Dropout 0.25 & - & - & Dense 2 Softmax & Dense 2 Softmax\\
\hline
\textbf{Layer 8} &  2 Dense Softmax & 2 Dense Softmax & - & - & - & - \\
\bottomrule[2pt]
\multicolumn{7}{l}{$\ast$ The models $f_{\ast}$ act as the defenders while $f'_{\ast}$ as attackers. The activation function of each layer is $ReLu$ unless specifically noted.} \\
\multicolumn{7}{l}{$\ast$ The kernel size is $3 \times 3$ for all the CNN models.}
\end{tabular}
\end{center}
\label{table:struct}
\end{table*}

\subsection{Metrics and Baselines}

\subsubsection{Metrics}

We set two metrics to evaluate the performance of adversarial attacks. Since all the test records are false measurements (generated by our random initialization scheme), the first metric is the detection recall (TP/(TP + FN)) of the defender's ML models under adversarial attacks. Meanwhile, it is straightforward that an adversarial measurement vector with a smaller profit to the energy thief will have a higher probability to bypass the detection. Therefore, we set the average $L_{1}$-Norm of the adversarial measurement vectors as the second evaluation metric. In our experiment, the average $L_{1}$-Norm of all normal measurement records is 32.05 kWh.

\subsubsection{Baselines}

We set up two \textbf{vanilla black-box attackers} as baselines. The first vanilla attacker $\textbf{VA1}$ will gradually try different $\alpha$ of $h_{1}$ as defined in Table \ref{table:scenario} while the second vanilla attacker $\textbf{VA2}$ generates uniformly distributed measurement vector between 0 and a variable $u$.

\subsection{Experimental Result}

The evaluation experiments are conducted with 1,000 adversarial measurement vectors. All the DNN's detection recall of the original randomly initialized adversarial measurement vectors is 100\%. The standard deviation of the Gaussian distribution used for initialization is set to $\sigma = 0.0001$. 

\subsubsection{Vanilla Attacks}

\begin{figure}[htbp]
\centerline{\includegraphics[width=1\linewidth]{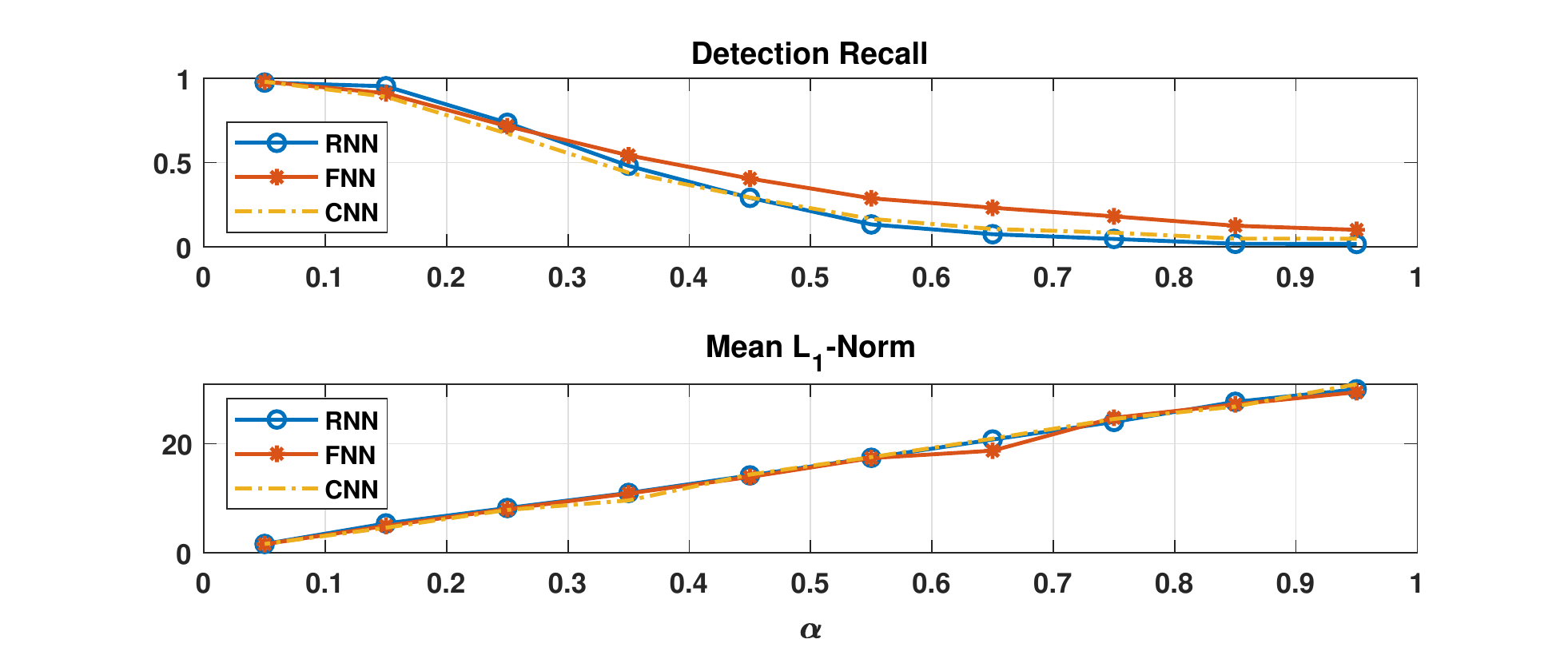}}
\caption{Vanilla attacker 1 evaluation result.}
\label{fig:vanilla1}
\end{figure}

As expected, the detection accuracy of the defenders' models decreases with the parameter $\alpha$ increases under $\textbf{VA1}$ attack. This indicates that $\textbf{VA1}$ has a higher success probability if he/she was willing to decrease his/her stolen profit. From Fig. \ref{fig:vanilla1}, if $\textbf{VA1}$ wants to have a relatively high success probability for energy theft, such as over 65\%, the required power consumption bill should be over 20 kWh ($\alpha > 0.65$).

\begin{figure}[htbp]
\centerline{\includegraphics[width=1\linewidth]{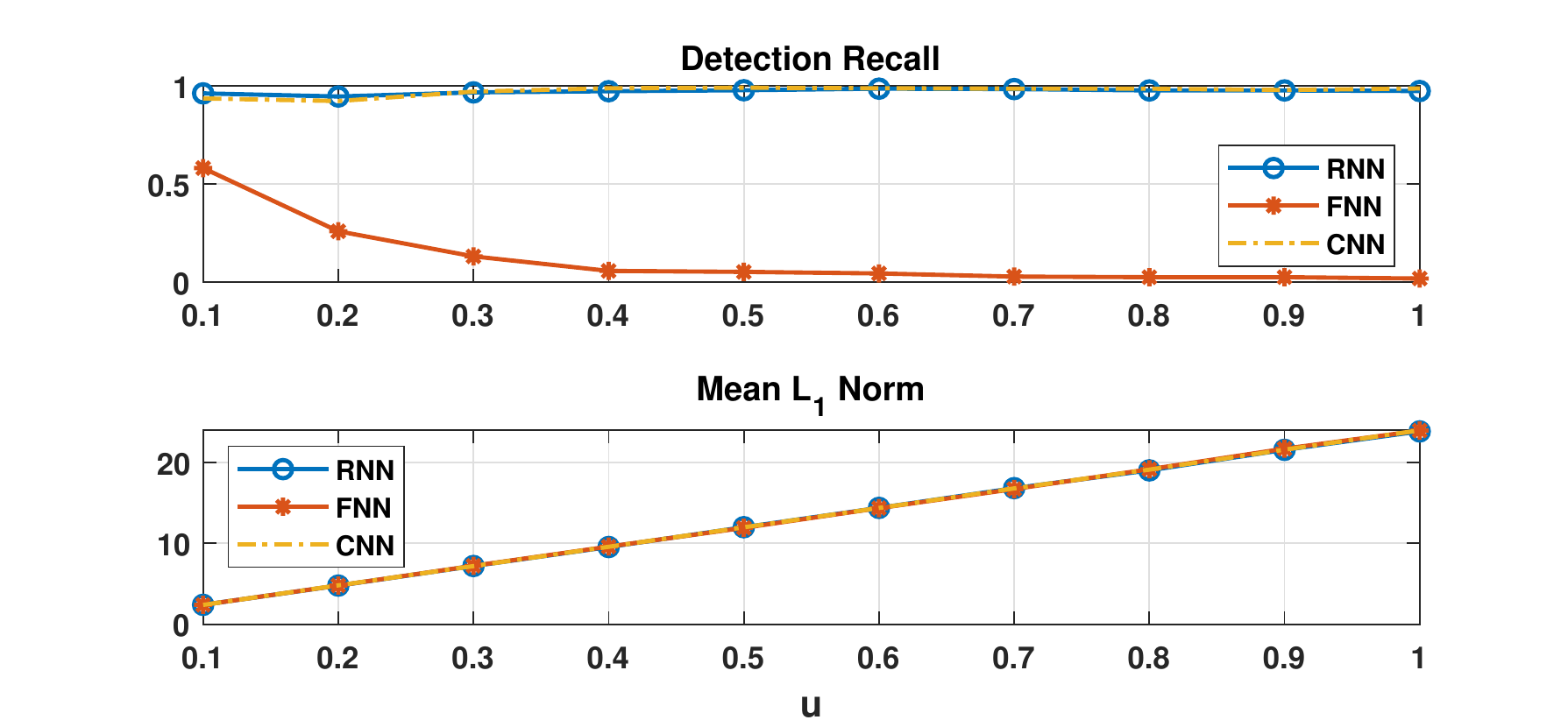}}
\caption{Vanilla attacker 2 evaluation result.}
\label{fig:vanilla2}
\end{figure}

As shown in Fig. \ref{fig:vanilla2}, the detection recall of RNN and CNN remains high (over 95\%) with the parameter $u$ increases. This indicates that a uniformly distributed consumption measurement vector is obviously abnormal for models that are trained to learn the daily electricity consumption patterns. Overall, the $\textbf{VA2}$ attack is not effective for energy theft for the attacker.

\subsubsection{State-of-the-art Approaches}

We apply the random initialization approach to the state-of-the-art AML algorithms discussed in Section \ref{sec:sotap} and evaluate the attack performances under the white-box and black-box settings. Similar to Algorithm \ref{al:ssf}, we map all the negative values in the adversarial measurements to zero to be feasible. We test different $\epsilon$ values for FGSM and FGV, and evaluation result is shown in Fig. \ref{eq:fgsm} and Fig. \ref{eq:fgv} respectively. 

\begin{figure}[htbp]
\centerline{\includegraphics[width=1\linewidth]{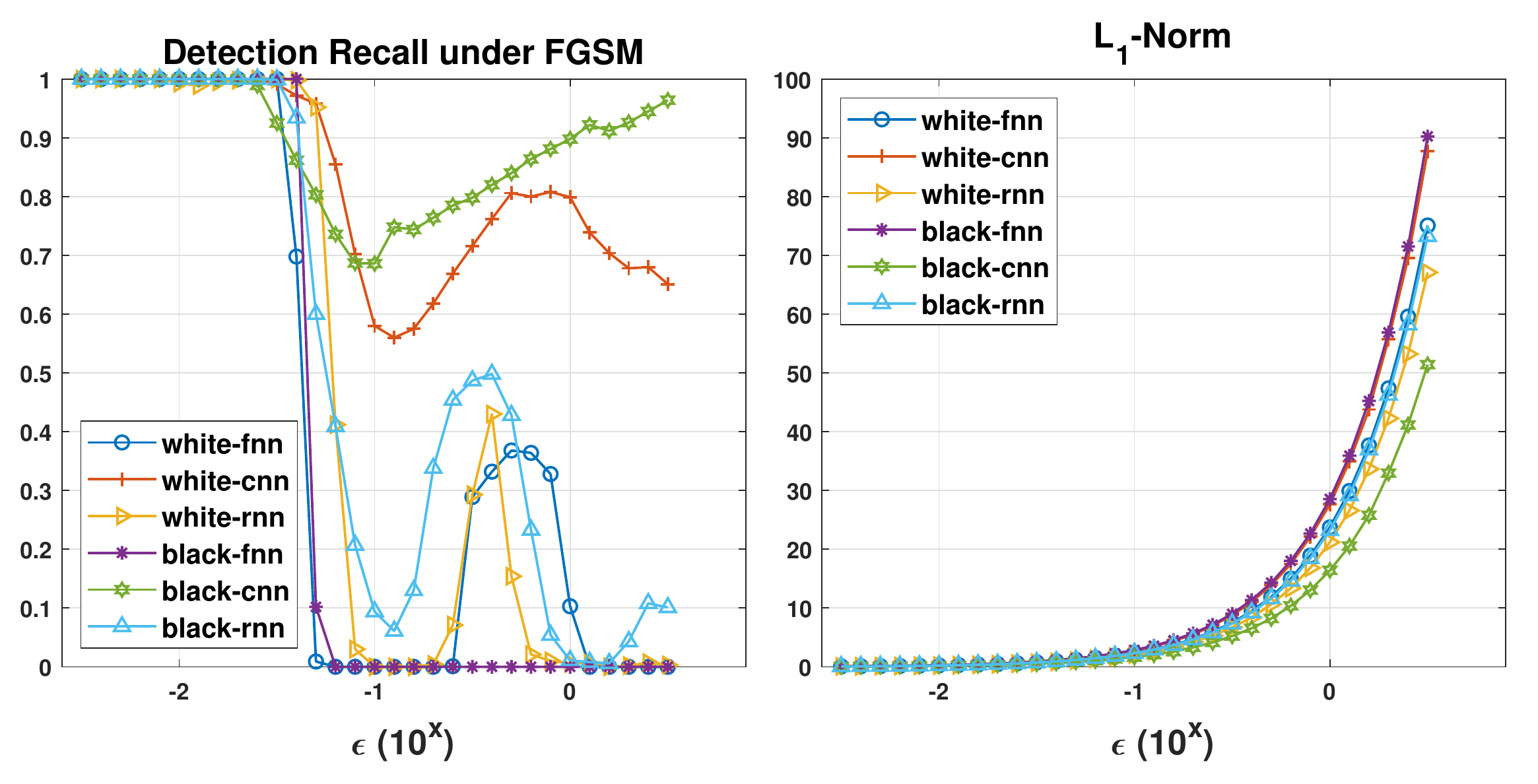}}
\caption{FGSM Evaluation Result}
\label{fig:fgsm}
\end{figure}

\begin{figure}[htbp]
\centerline{\includegraphics[width=1\linewidth]{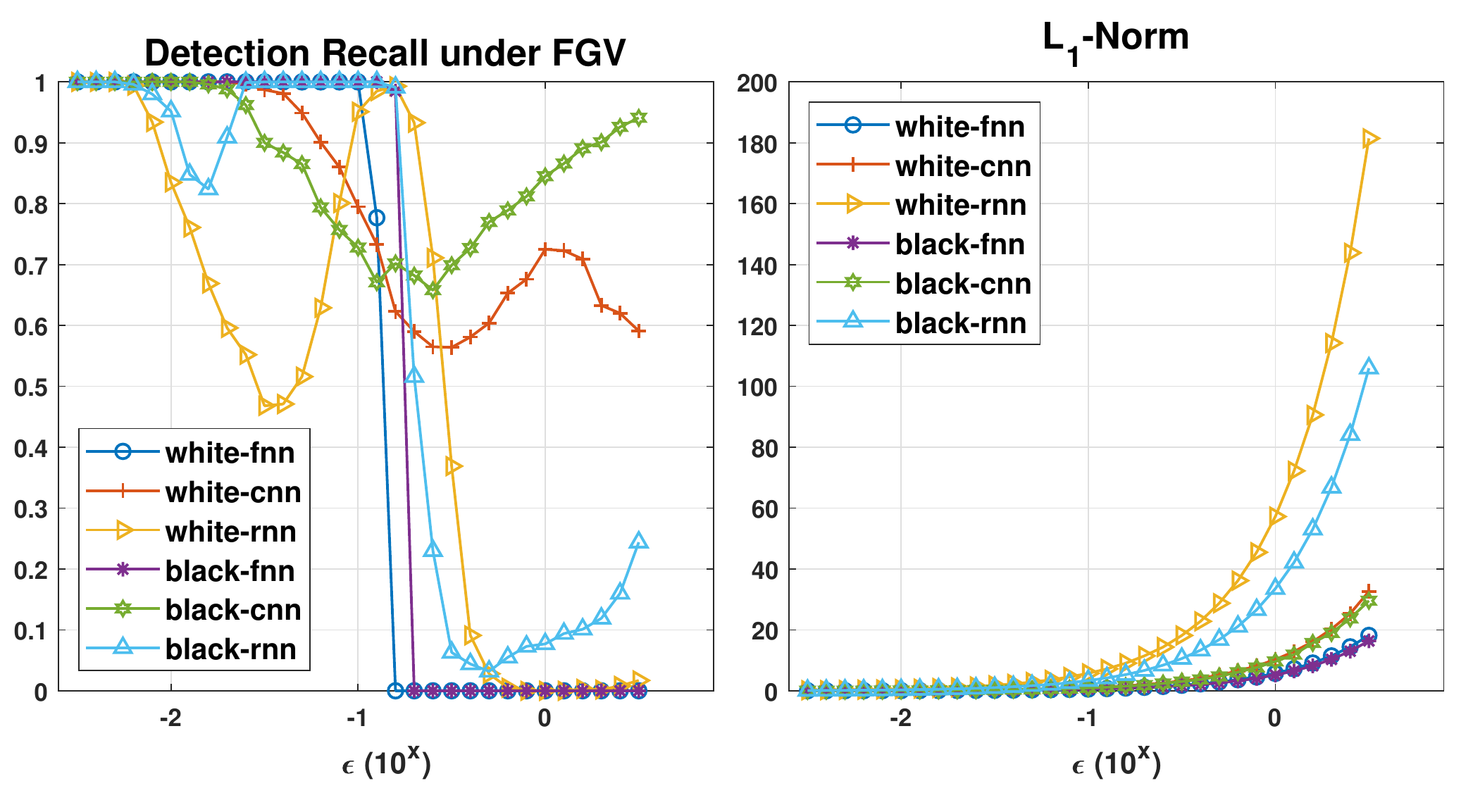}}
\caption{FGV Evaluation Result}
\label{fig:fgv}
\end{figure}

From Fig.~\ref{fig:fgsm}, we can learn that FGSM can achieve notable attack performance for FNN and RNN. In the black-box settings, the probability of bypassing RNN detection is over 90\% while the adversarial measurement's $L_1$-Norm is only 2.9 kWh ($\epsilon = 10^{-0.9}$). The attack performance is even better for FNN. When $\epsilon = 10^{-1.2}$, the energy thief obtains a 100\% detection bypassing rate with a 1.4 kWh electricity bill. The single-step attack FGSM does not perform well for CNN detection. The best evasion rate is around 44\% for the white-box attack and 32\% for the black-box attack.

Overall, the attack performance of FGV is slightly worse than FGSM in black-box settings but is still effective, as shown in Fig.~\ref{fig:fgv}. For example, the black-box attack to RNN obtains a 94\% detection bypassing rate while the $L_1$-Norm is 10.6 kWh ($\epsilon = 10^{-0.5}$), which is higher than FGSM (2.9 kWh) but is still smaller than the normal measurements (32.05 kWh). Similar to FGSM, the FGV achieves the best performance for FNN detection, followed by RNN and CNN.

\begin{table}[htbp]
\caption{DeepFool Evaluation Performance}
\begin{center}
\begin{tabular}{c|c|c|c|c|c|c}
\toprule[2pt]
\textbf{Model} & \multicolumn{2}{|c|}{\textbf{FNN}} & \multicolumn{2}{|c|}{\textbf{CNN}} & \multicolumn{2}{|c}{\textbf{RNN}}\\
\hline
\textbf{Metric} & recall & size & recall & size & recall & size \\
\hline
\textbf{white-box} & 0\% & 0.94 & 0\% & 0.23 & 0\% & 0.02\\
\hline
\textbf{black-box} & 17.4\% & 1.14 & 100\% & 0.115 & 100\% & 0.06 \\
\bottomrule[2pt]
\multicolumn{7}{l}{$\ast$ `size' is the $L_1$-Norm of adversarial measurements (kWh)} \\
\end{tabular}
\end{center}
\label{table:deepfool}
\end{table}

The evaluation results of the iterative attack DeepFool is summarized in Table \ref{table:deepfool}. The iterative attack demonstrates notable performances in white-box settings. The detection recall of all three DNNs becomes 0\% under white-box attacks while the $L_1$-Norm is smaller than 1 kWh. However, as expected, the adversarial measurements generated by iterative attacks are less transferable. Under the black-box setting, the DeepFool attack only shows effectiveness in FNN detection while the detection recall of CNN and RNN remains 100\%.

\subsubsection{SearchFromFree Iteration Algorithm}

\begin{figure}[htbp]
\centerline{\includegraphics[width=1\linewidth]{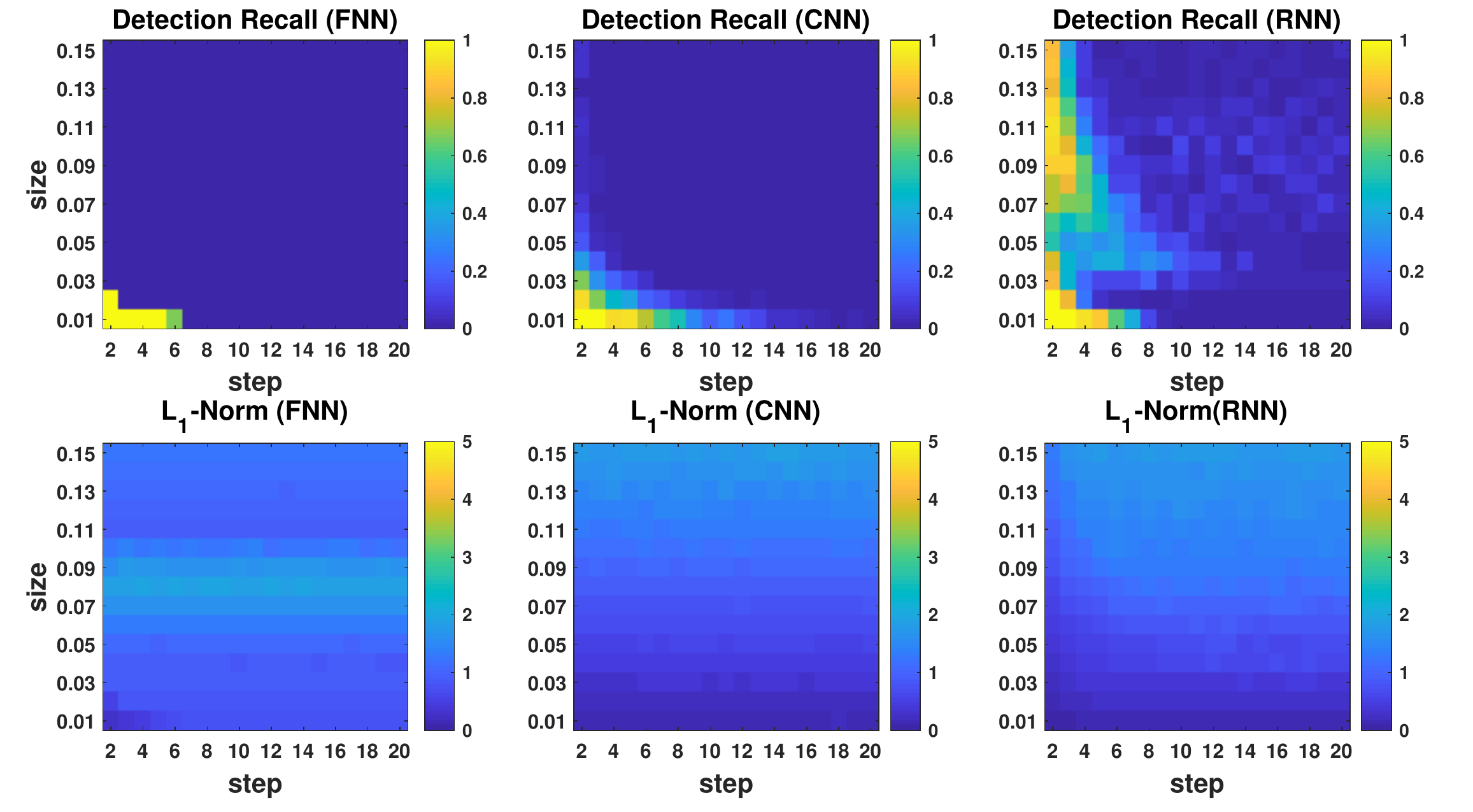}}
\caption{The evaluation performance of SearchFromFree Iteration algorithm (white-box attacks).}
\label{fig:ssfwhite}
\end{figure}

\begin{figure}[htbp]
\centerline{\includegraphics[width=1\linewidth]{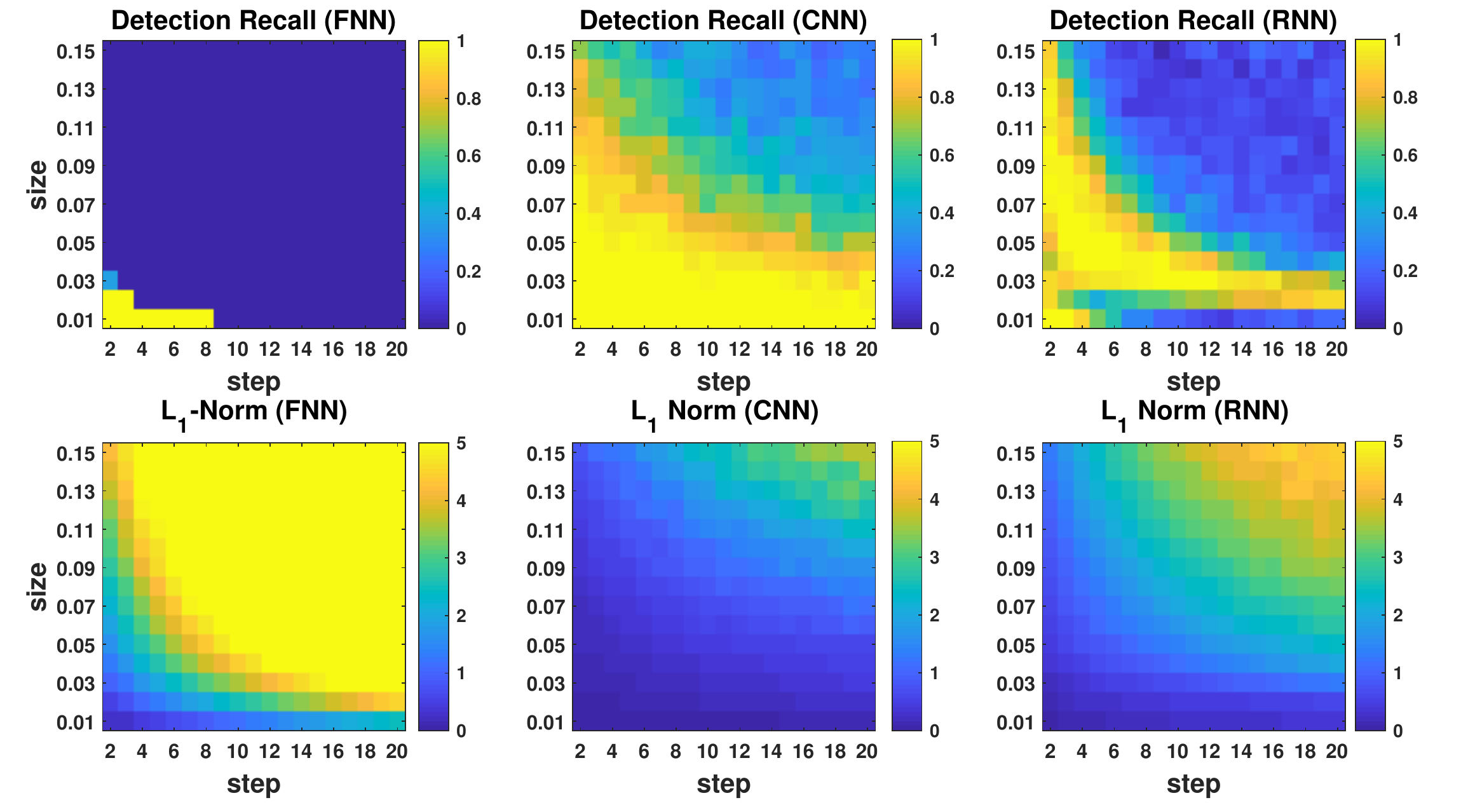}}
\caption{The evaluation performance of SearchFromFree Iteration algorithm (black-box attacks).}
\label{fig:ssfblack}
\end{figure}

We then evaluate the performance of our \textbf{ssf-Iter} algorithm, an iterative attack algorithm that utilizes a step-size scheme to search for transferable adversarial measurements, as shown in Fig.~\ref{fig:ssfwhite} and Fig.~\ref{fig:ssfblack}.

From Fig. \ref{fig:ssfwhite}, we can learn that our algorithm performs best in FNN, followed by CNN and RNN. In most cases, the detection recall of three DNNs approaches to zero under the white-box attack while the adversarial measurements are still small enough (around 1 kWh). 

As expected, the adversarial performances under the black-box setting are worse than the white-box setting, as shown in Fig. \ref{fig:ssfblack}. In general, the probability of bypassing the detection is lower and the corresponding measurement size is larger. Meanwhile, the attacker is required to pay a higher cost ($L_1$-Norm of adversarial measurements) in the black-box settings to obtain the same detection bypassing rate in the white-box settings. Statistically, the FNN detection still performs worst under black-box adversarial attacks. By analyzing the corresponding evaluation parameters, we can learn that the attacker can bypass the FNN's detection with nearly 100\% success probability while the average $L_1$-Norm is around 1 kWh. For CNN detection, our adversarial attack can achieve over 70\% successful rate while keeping the $L_1$-Norm below 4 kWh. Attack performance is better for RNN detection. In most attack scenarios, the RNN's detection recall is below 30\% while the $L_1$-Norm is lower than 3 kWh. It is worth noting that if the attacker sets $size$ to 0.01, the adversarial attack can obtain over 80\% successful probability with an around 0.2 kWh measurement size. Compared with the DeepFool attack, our algorithm achieves similar performance in the white-box settings and better transferability under the black-box settings.

\noindent \textbf{Parameter Selection: } Fig. \ref{fig:ssfwhite} and Fig. \ref{fig:ssfblack} show that the attack performances can be impacted by the parameters in Algorithm \ref{al:ssf}. However, from the 2D pixel figures, we can observe that the attack performances follow specific patterns according to the two parameters. Overall, as long as the parameters fall in a specific range, the attack performance will be satisfied. Meanwhile, by comparing Fig. \ref{fig:ssfwhite} and Fig. \ref{fig:ssfblack}, we can learn that the performances of black-box attacks share similar manifolds with white-box attacks under our step-size scheme. This indicates that the attacker can select the algorithm parameters based on the performances of his/her local DL models. In practical scenarios, different attackers may also collude together to search for the parameters that produce the best attack performance.

\section{Adversarial Defense} \label{sec:defense}

\subsection{Adversarial Defense Analysis}

Section~\ref{sec:evaluation} demonstrates that the low-cost adversarial measurements (below 1 kWh) can bypass the DL detection with a high probability (overall 70\%), which indicates that the DL models have large input space that the attacker can take advantage of in the DL detection model. For example, the RNN curve in Fig.~\ref{fig:fgsm} cannot detect adversarial measurements that fall into specific ranges. Enabled by different AML algorithms, the attacker can finally find such blind spots of the DL models. In fact, adversarial examples will always exist as long as the DL model is not perfect since the attacker can always modify the input to change the model's prediction result. For example, if the vanilla attacker 1 sets the parameter $\alpha$ to a large value, such as $\alpha = 1-10^{-5}$, the adversarial vectors will hold a very high probability to be classified as normal.

Although adversarial examples are inevitable, corresponding defense mechanisms can still be employed to make adversarial attacks infeasible in practice or increase the attacker's effort to launch attacks. In the CV domain, if the attacker generates a large perturbation to deceive the DL model, the adversarial images will be abnormal for human eyes, and the attack performance decreases. Accordingly, in energy theft detection, the objective of the adversarial defense should be increasing the cost ($L_1$-Norm of the adversarial measurement vectors) the attacker needs to pay to bypass the DL detection.

\begin{figure}[htbp]
\centerline{\includegraphics[width=0.95\linewidth]{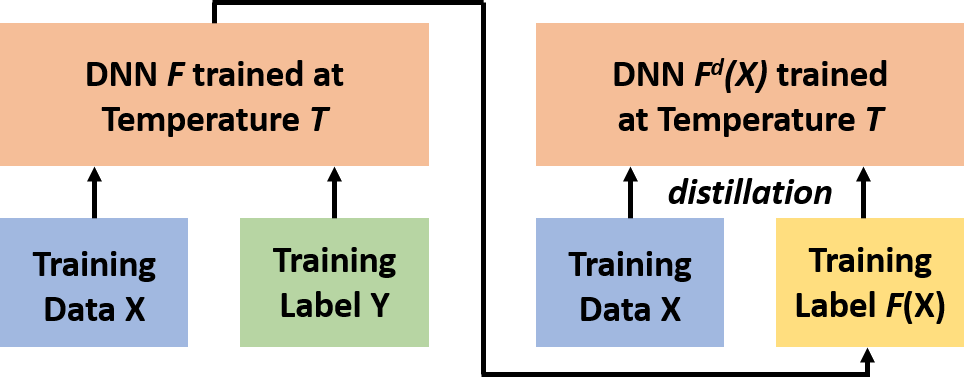}}
\caption{Overall structure of model distillation \cite{papernot2016distillation}.}
\label{fig:distillation}
\end{figure}

Currently, the main state-of-the-art defense approaches in the CV domain includes adversarial training \cite{kurakin2016adversarial}, model distillation \cite{papernot2016distillation}, adversarial detection \cite{ma2019nic} and input reconstruction \cite{meng2017magnet}. \cite{kurakin2016adversarial} has shown that adversarial training is effective for single-step attacks but performs deficiently for iterative attacks. Adversarial detection and input reconstruction assume that the adversarial examples follow different manifolds from normal examples and detect adversarial examples before feeding them to the victim model. However, they are shown to be vulnerable to dynamic and iterative attacks \cite{meng2017magnet, metzen2017detecting, gu2014towards}. In this paper, we study the defense performance of model distillation and demonstrate it can mitigate the threat of adversarial attacks in energy theft detection.

\subsection{Model Distillation Defense}

The basic principle of model distillation is demonstrated in Fig.~\ref{fig:distillation}. A DNN $F$ is first trained with the original dataset $\left \{ X, Y \right \}$. The prediction probability vector $F(X)$ from the last softmax layer of $F$ is then used as the label to train the distilled model $F^{d}$ that owns the same structure with $F$. \cite{papernot2016distillation} demonstrates that the distilled model $F^{d}$ becomes less sensitive to input perturbation and is more robust to adversarial examples.

\begin{table}[htbp]
\caption{Distilled DNN Performance}
\begin{center}
\begin{tabular}{c|c|c|c|c|c|c}
\toprule[2pt]
\textbf{Model} & \multicolumn{2}{|c|}{$f^{d}_{FNN}$} & \multicolumn{2}{|c|}{$f^{d}_{CNN}$} & \multicolumn{2}{|c}{$f^{d}_{RNN}$}\\
\hline
\textbf{Metric} & Accu & FPR & Accu & FPR & Accu & FPR \\
\hline
\textbf{Performance} & 83.6\% & 16.3\% & 95.17\% & 5.2\% & 97.7\% & 2.5\%\\
\bottomrule[2pt]
\multicolumn{7}{l}{$\ast$ `Accu' is the overall classification accuracy} \\
\multicolumn{7}{l}{$\ast$ `FPR' is the false positive rate} \\
\end{tabular}
\end{center}
\label{table:dis}
\end{table}

\begin{figure}[htbp]
\centerline{\includegraphics[width=1\linewidth]{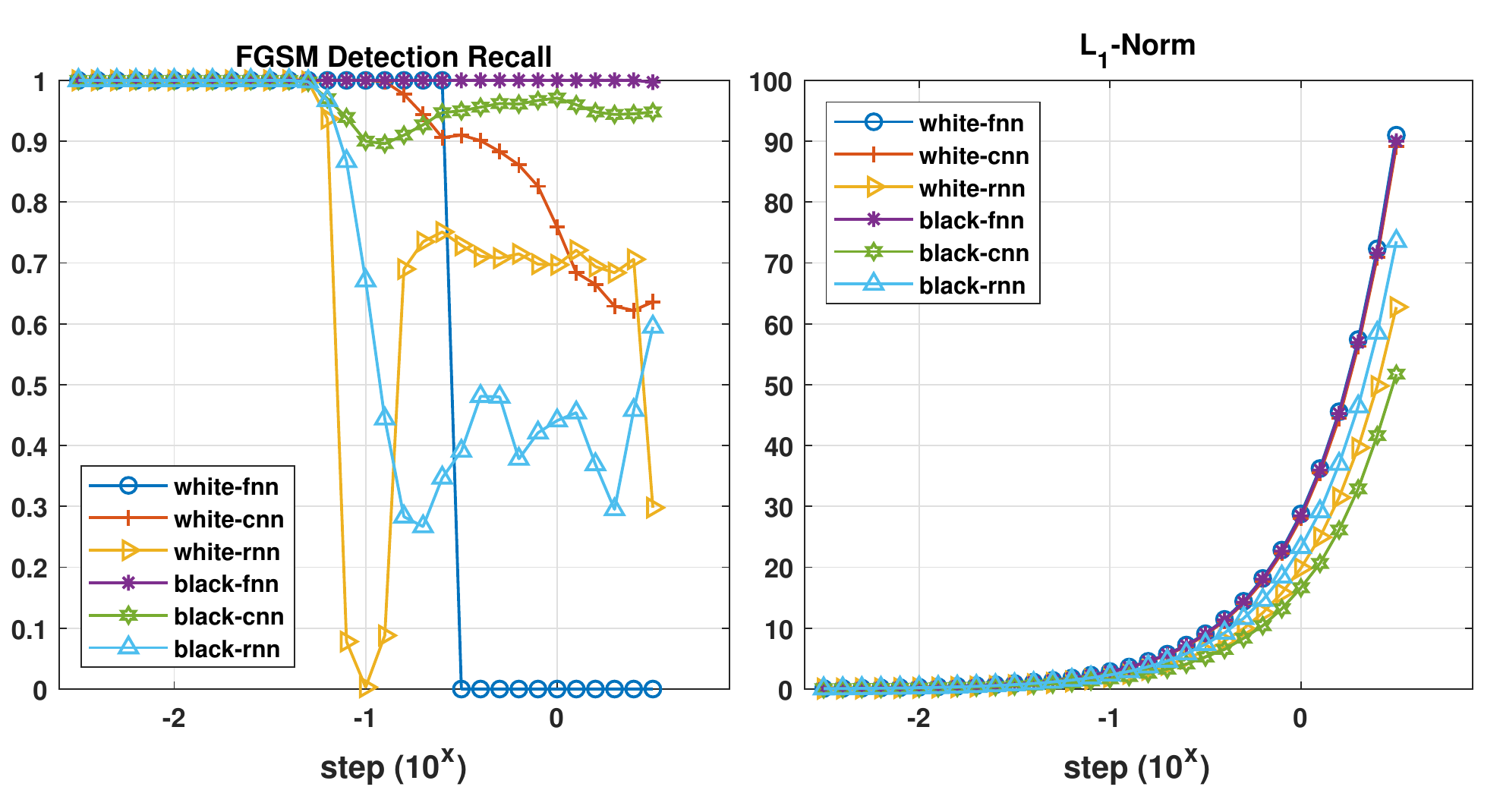}}
\caption{FGSM evaluation with model distillation. ($T=100$)}
\label{fig:fgsm-dis}
\end{figure}

\begin{figure}[htbp]
\centerline{\includegraphics[width=1\linewidth]{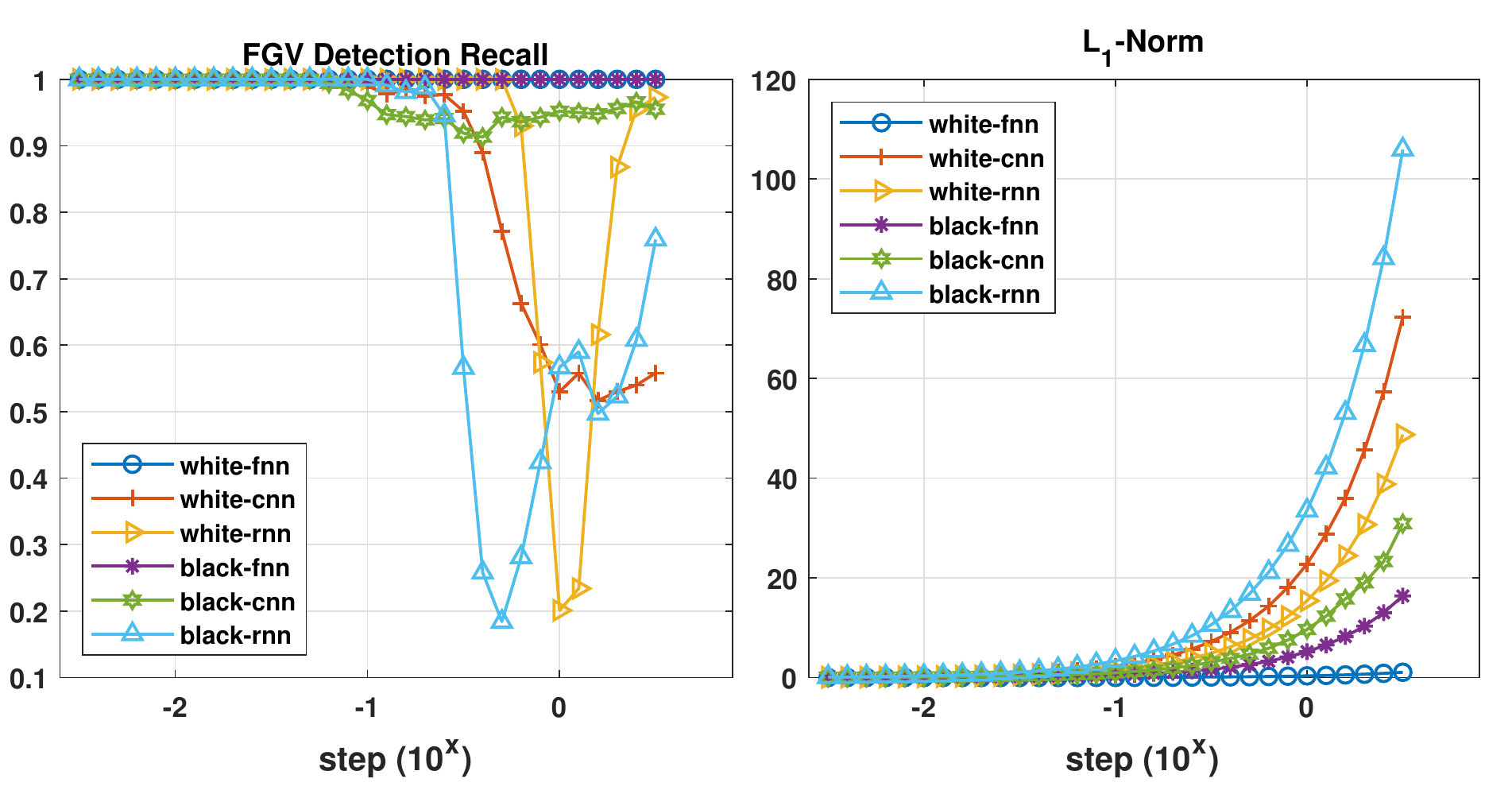}}
\caption{FGV evaluation with model distillation. ($T=100$)}
\label{fig:fgv-dis}
\end{figure}

We train three distilled DNNs for the utilities with the corresponding structures described in Table \ref{table:struct}. We set the distillation temperature to $T = 100$ and the overall detection performances are shown in Table \ref{table:dis}. We then launch FGSM and FGV attacks to the distilled DNNS and the evaluation results are shown in Fig.~\ref{fig:fgsm-dis} and Fig.~\ref{fig:fgv-dis} respectively.

By comparing Fig.~\ref{fig:fgsm-dis} and Fig.~\ref{fig:fgsm}, we can observe that all the distilled models demonstrate higher reliability against the black-box FGSM attack, especially FNN and CNN. For the distilled RNN, the best detection bypassing rate is 72\% with the $L_1$-Norm equals 4.6 kWh, which are still worse than the plain RNN (94\% and 3.6 kWh). The distilled approach is also effective for the FGV attack. As shown in  Fig.~\ref{fig:fgv-dis} and Fig.~\ref{fig:fgv}, the model distillation significantly increases the reliability of FNN and CNN under black-box attacks (0\% and 9\% bypassing rate). The black-box FGV attack can still achieve a 44\& bypassing rate with an 11 kWh $L_1$-Norm for the distilled RNN but is obviously worse than the plain RNN (77\% and 8.4 kWh).

We also evaluate the defense performances of distilled DNNs on mitigating our black-box iterative attacks, as shown in Fig.~\ref{fig:ssfblack}. Overall, the black-box attacker is required to pay more cost to obtain similar detection bypassing performance, especially for FNN, and CNN. However, the attacker can still achieve an overall 70\% detection bypassing rate with a below 4 kWh energy cost for RNN detection.

\begin{figure}[htbp]
\centerline{\includegraphics[width=1\linewidth]{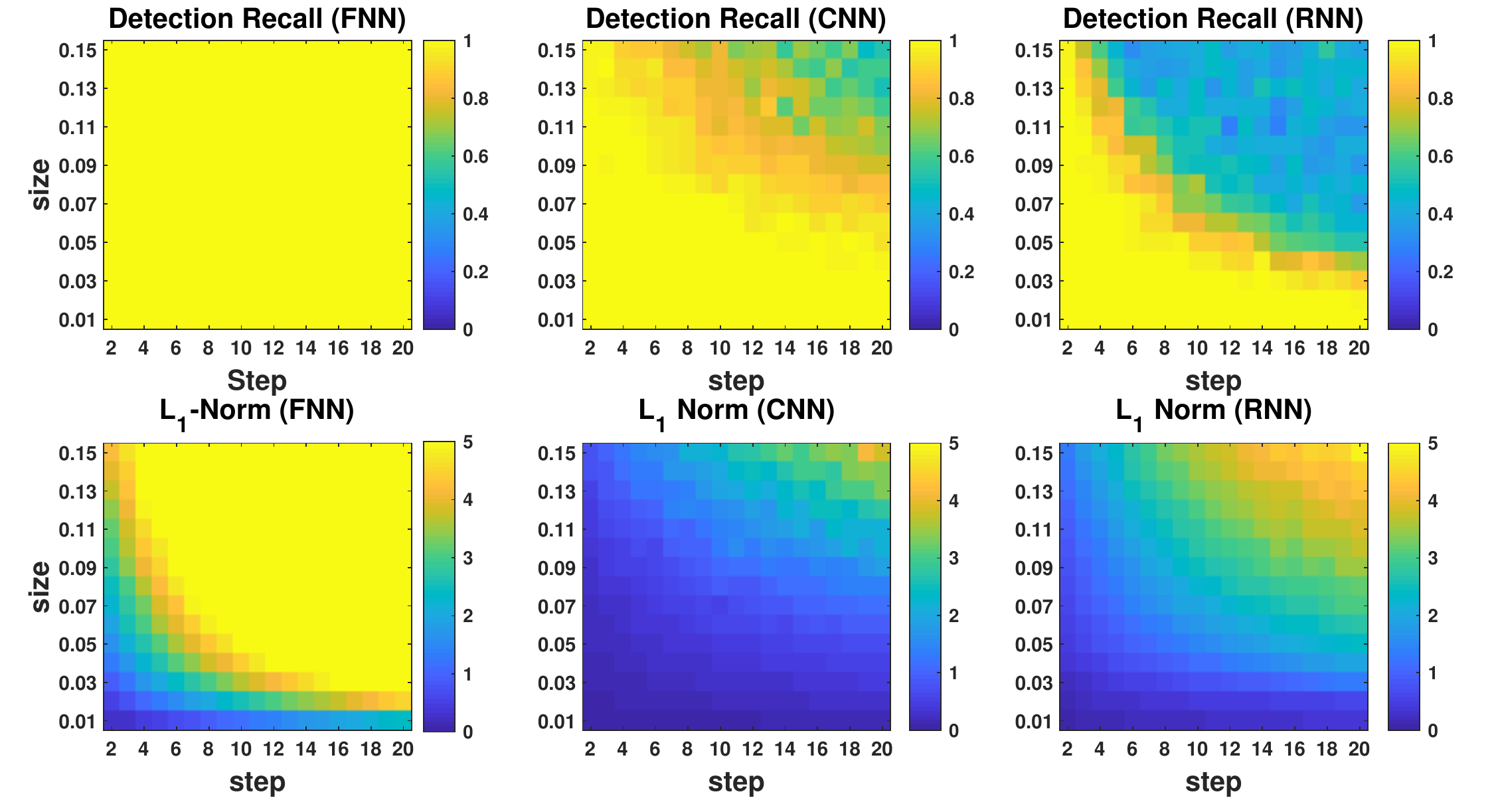}}
\caption{Black-box SearchFromFree Iteration evaluation with model distillation. ($T=100$)}
\label{fig:ssf-black-dis}
\end{figure}

\section{Discussion and Conclusion} \label{sec:conclusion}

In this paper, we demonstrate that the well-performed DL models proposed in the literature are highly vulnerable to adversarial attacks. We design a random initialization approach that is compatible with different AML algorithms to increase the energy thief's stolen profit. Meanwhile,  a step-size iteration scheme is proposed to increase the transferability of black-box iterative attacks. The evaluation results show that our framework enables the energy thief to report extremely low power consumption measurement data to the utilities without being detected by the DL models. 

We also discuss the corresponding defense mechanisms of adversarial attacks in energy theft detection and analyze the defense requirements. In this work, we implement the model distillation scheme and show that it can mitigate the adversarial attacks and increase the attacker's cost. However, the attacker is still able to achieve considerable attack performance under black-box settings. In the future, practical defensive approaches that can effectively increase the DNN's robustness against adversarial examples will be studied, such as joint detection.

\bibliography{reference} 
\bibliographystyle{ieeetr}

\end{document}